\documentclass[a4paper,10pt,pra,twocolumn,superscriptaddress,floatfix]{revtex4-2}

\pdfoutput=1
\usepackage[utf8]{inputenc}
\usepackage{amsmath}
\usepackage{hyperref}
\usepackage{dsfont}
\usepackage{xspace}
\usepackage{graphicx}
\usepackage{subfigure}

\usepackage{tikz}
\usetikzlibrary{quantikz}
\usetikzlibrary{shapes}
\usetikzlibrary{arrows}
\usetikzlibrary{shadows}
\usetikzlibrary{calc}
\usetikzlibrary{positioning}

\newsavebox\CPU
\sbox\CPU{
    \begin{tikzpicture}[line cap=round,line join=round,scale=1]

    \draw[thick,rounded corners,fill=white] (0,0) rectangle ++ (1.2,1.2);
    \draw[thick,rounded corners] (0.15,0.15) rectangle ++ (0.9,0.9);
    \draw[draw,thick] (0.6,0) -- (0.6,-0.2);
    \draw[draw,thick] (0.4,0) -- (0.4,-0.2);
    \draw[draw,thick] (0.8,0) -- (0.8,-0.2);

    \draw[draw,thick] (0.6,1.2) -- (0.6,1.4);
    \draw[draw,thick] (0.4,1.2) -- (0.4,1.4);
    \draw[draw,thick] (0.8,1.2) -- (0.8,1.4);

    \draw[draw,thick] (0,0.4) -- (-0.2,0.4);
    \draw[draw,thick] (0,0.6) -- (-0.2,0.6);
    \draw[draw,thick] (0,0.8) -- (-0.2,0.8);

    \draw[draw,thick] (1.2,0.4) -- (1.4,0.4);
    \draw[draw,thick] (1.2,0.6) -- (1.4,0.6);
    \draw[draw,thick] (1.2,0.8) -- (1.4,0.8);

    \node[] at (0.6,0.6) {\small{\textsf{CPU}}};
    \end{tikzpicture}}

\tikzstyle{qdev} = [draw, rectangle, 
    minimum height=3em, fill=none, minimum width=2.5em, rounded corners, node distance=1cm]

\tikzstyle{anch} = [fill=none, minimum width=3em, minimum height=0.1cm, node distance=1cm]

\tikzstyle{jam1} = [draw, rectangle, dashed, line width=0.5pt, fill=blue!35,
    minimum width=2.5em, minimum height=1.25cm, rounded corners]

\tikzstyle{jam2} = [draw, rectangle, dashed, line width=0.5pt, fill=blue!35,
    minimum width=3em, minimum height=3cm, rounded corners]
    
\tikzstyle{jam3} = [draw, rectangle, dashed, line width=0.5pt, fill=blue!35,  minimum width=3em, minimum height=4.5cm, rounded corners]

\tikzstyle{ssfb} = [draw, rectangle, dotted, line width=0.75pt, fill=black!10, minimum width=3em, minimum height=3cm, rounded corners]

\tikzstyle{vqfe} = [draw, rectangle, dashdotted, line width=0.5pt, fill=yellow!75, minimum width=3em, minimum height=1in, rounded corners]

\tikzstyle{vqfe-short} = [draw, rectangle, dashdotted, line width = 0.5pt,    fill=yellow!75, minimum width=3em, minimum height=0.5in, rounded corners]

\tikzstyle{cpu} = [node contents=\usebox{\CPU},scale=0.75,]

\newcommand{\ketbra}[2]{\ensuremath{|#1 \rangle\!\langle #2|}}
\newcommand{\tr}{\ensuremath{\mathrm{tr}}}

\newcommand{\id}{\mathds{1}}
\newcommand{\Id}{\id}

\newcommand{\Cplx}{\ensuremath{\mathds{C}}}

\newcommand{\ie}{\emph{ie.}}

\newcommand{\Jam}[1]{\ensuremath{\mathcal{J}(#1)}}

\newcommand{\figref}[1]{Fig.~(\ref{#1})}

\begin{document}
\title{Variational certification of quantum devices}

\author{Akash Kundu}
\affiliation{Institute of Theoretical and Applied Informatics, Polish Academy of Sciences, Ba{\l}tycka 5, 44-100 Gliwice, Poland}

\affiliation{Joint Doctoral School, Silesian University of Technology, Akademicka 2A, 44-100 Gliwice, Poland}
\email{akundu@iitis.pl}

\author{Jaros{\l}aw Adam Miszczak}
\affiliation{Institute of Theoretical and Applied Informatics, Polish Academy of Sciences, Ba{\l}tycka 5, 44-100 Gliwice, Poland}
\email{jmiszczak@iitis.pl}


\begin{abstract}
One of the requirements imposed on the realistic quantum computers is to provide computation results which can be repeated and reproduced. In the situation when one needs to repeat the quantum computation procedure several times, it is crucial that the copies of the quantum devices are similar in the sense of the produced results. In this work, we describe a simple procedure based on variational quantum eigensolver which can be utilized to compare quantum devices. The procedure is developed by combining Choi-Jamio{\l}kowski isomorphism with the variational hybrid quantum-classical procedure for matrix diagonalization. We compare the introduced procedure with the scheme based on the standard bounds for the similarity between quantum operations by analysing its action on random quantum channels. We also discuss the sensitivity of the described procedure to the noise, and we provide numerical results demonstrating its feasibility in realistic scenarios by running the procedure on IBM quantum computer.

\vspace*{6pt}
\end{abstract}

\maketitle
\section{Introduction}
Quantum information processing aims at utilizing the rules of quantum mechanics for the purpose of transmitting information~\cite{ingarden1976quantum} and processing data~\cite{deutsch1985quantum}. To this end one needs to construct quantum mechanical devices which implement the desired protocols or algorithms. Moreover, to make any such device suitable for deployment in real-world applications one needs to make sure that the fabricated system indeed executes the desired quantum operation and prepares the required quantum states.

At the moment, among the most promising approaches for providing useful quantum computers one can point the development of quantum procedures based on small and limited devices. Such devices are commonly referred to as NISQ (noisy intermediate scale quantum) computers because they should be available in the foreseeable future, will be rather small, limited by the number of qubits and the available set of gates, and their operational capabilities will be severely limited by errors~\cite{preskill2018quantum}.

Even for NISQ-era devices, working in the $10^2$ qubits regime, to ensure that the quantum device provided by the merchant or the developer indeed executes the desired task, it must undergo the certification procedure.
Even if the accurate control of quantum information systems foreshadows numerous applications in the area of data processing, the complexity of such devices renders the certification of their correct operating a significant challenge~\cite{eisert2019quantum,kliesch2020theory}.

Certification tasks are particularly demanding in quantum computing applications as the task of certifying the properties of a quantum system is equivalent to reproducing the result obtained by it in the classical simulation. This, however, is computationally hard, which is exactly the point of the quantum supremacy~\cite{harrow2017quantum,boixo2018characterizing,arute2019quantum}.

The difficulty of the certification of the quantum devices stems mainly from the intrinsic computational advantages offered by quantum computers. Thus, it is natural to consider the possibility of utilizing quantum computers for providing the certification of quantum devices. In this work we investigate such a scenario by providing a certification scheme based on the structure of the space of quantum operations. The natural correspondence between states and operations in the quantum mechanics gives rise to new methods of information processing beyond the scope of classical mechanics.

The ideas presented in this work are organized according to the following plan. 
We start with the preliminaries concerning the problems related to the distinguishability of states and operations, and we recall standard mathematical apparatus utilized for investigating these matters.
Next, in Section~\ref{sec:fidelity-estimation}, we consider a scenario in which quantum devices need to undergo a certification procedure and introduce two schemes. In Section~\ref{sec:proc-compare}, we compare the discussed certification procedures and investigate their behaviour for the random quantum operations. We also provide the analysis of the impact of noise models on the efficiency of the variational fidelity estimation for random quantum operations.  We conclude the paper in Section \ref{sec:final} by discussing the features and limitations of the introduced schemes and their possible applications.

\section{Preliminaries}

One of the most prominent characteristics of quantum information is its susceptibility to errors. Even if theoretical considerations and schemes are based on the assumption that quantum states are represented by normalized vectors, in most cases this is not true when one needs to deal with the actual physical realization of the quantum computation processes. It is clear that the  errors occurring during the process of preparation and processing of data encoded in a quantum system are unavoidable and have to be taken into account during the development of quantum procedures for realistic applications. From the mathematical point of view this means that the state of a quantum system during the process of quantum computation has to be described by s density matrice which is represented by the convex combinations of pure states~\cite{holevo2001statistical,keyl2002fundamentals}.

From the theoretical point of view, the problem of distinguishing between quantum devices is identical to the problem of defining the measure of similarity on the space of density matrices.
The interest in the physical realizations of quantum information processing resulted in considerable effort in this area~\cite{ji2006identification,wang2006unambiguous,piani2009all,duan2009perfect}, with the special focus  on the feasibility of the proposed schemes \cite{sedlak2009unambiguous}.
What's more, the same mathematical apparatus developed for solving discrimination problems can be used to quantify the amount of errors in physical realizations of quantum processes~\cite{gilchrist2005distance}. This issue is crucial for describing the precision of the physical implementation of quantum information processing.

The most common measure of similarity between quantum states $\rho$ and $\sigma$ is given by fidelity~\cite{uhlmann1976transition}, 
\begin{equation}
F(\rho,\sigma) = \|\sqrt{\rho}\sqrt{\sigma}\|_1 = \tr\sqrt{\sqrt{\rho}\sigma\sqrt{\rho}},
\end{equation}
where $\|\cdot\|_1$ is the Schatten 1-norm. Fidelity provides a quantum counterpart of the Bhattacharyya coefficient measuring the similarity between two probability distributions and it reduces to the scalar product for rank-1 operators. The important characteristic of fidelity is that its calculation requires non-integer powers of $\rho$ and $\sigma$ and thus there is no exact quantum algorithm which could be used to calculate it using finite number of copies of the input states.

As it is believed that computing quantum fidelity will be important to verify and characterize the states prepared on a quantum computer, a significant research effort has been devoted to the problem of finding methods for calculating or approximating it~\cite{liang2019quantum}. Therefore, it is also reasonable to provide the bounds for the values of fidelity using the functionals which could be easily calculated.
To address this, sub- and super-fidelity bounds (SSFB) for the fidelity between quantum states were introduced \cite{miszczak2009sub-super}, defined respectively as
\begin{equation}
E(\rho,\sigma) = \tr\rho\sigma + \sqrt{2[(\tr\rho\sigma)^2-\tr(\rho\sigma)^2 ]}
\end{equation}
and
\begin{equation}
G(\rho,\sigma) = \tr\rho\sigma + \sqrt{(1-\tr\rho^2)(1-\tr\sigma^2)},
\end{equation}
and fulfilling the property
\begin{equation}
E(\rho,\sigma) \leq F(\rho,\sigma) \leq G(\rho,\sigma).
\end{equation}

Both sub- and super-fidelity display many features desirable from the quantity used to approximate the fidelity. First of all, both are expressed in the form of functional of density matrices which have a direct representation in the form of quantum circuits~\cite{ekert2002direct}. In the case of two-qubit photonic states these quantities can be efficiently measured \cite{bartkiewicz2013direct}. Moreover, super-fidelity can be used to provide the lower bound for the trace distance between two quantum states~\cite{puchala2009bound}, that can be estimated using a simple measurement procedure. 

More recently, an alternative approach for computing the bounds for  fidelity, based on variational hybrid quantum-classical algorithm  has been proposed \cite{cerezo2020variational}. This variational quantum fidelity estimation (VQFE) algorithm was designed as the extension of the variational diagonalization procedure introduced in \cite{larose2019variational}. The variational diagonalization procedure enables finding $m$ eigenvalues of one of density matrices used to calculate fidelity in the eigenbasis of the second density matrix. Thus, the series of approximations -- so-called \emph{fidelity spectrum} -- can be obtained and used to bound the value of  fidelity. The bounds on the fidelity between states $\rho$ and $\sigma$ are expressed by
\begin{equation}\label{eq:truncated-fidelity-bounds}
F(\rho_m, \sigma_m^\rho) \leq F(\rho, \sigma) \leq F_\star(\rho_m^\rho, \sigma_m^\rho),
\end{equation}
with
\begin{equation}
F_\star(\rho_m, \sigma_m^\rho)\equiv \|\sqrt{\rho^\rho_m}\sqrt{\sigma_m^\rho}\|_1 + \sqrt{(1-\tr\rho^\rho_m)(1-\tr\sigma^\rho_m)},
\end{equation}
where $\sigma_m^\rho =\Pi_m^\rho \sigma \Pi_m^\rho$ denotes the operator obtained as the projection of $\sigma$ onto the subspace spanned by $m$ eigenvectors of $\rho$, corresponding to the largest eigenvalues and $\Pi_m^\rho$ is the projector onto this subspace. Quantity $F(\rho_m, \sigma_m^\rho)$ is called \emph{truncated fidelity}, and it has also been applied to the problem of computing quantum Fisher information~\cite{sone2021generalized}.

The VQFE algorithm targets the situations when one of the input states is low-rank. However, even if the procedure was developed by assuming low-rank approximations of a density matrix, the bounds given by \eqref{eq:truncated-fidelity-bounds} can be arbitrarily refined by increasing $m$.

In what follows, we utilize the SSFB and VQFE procedures as building blocks for the quantum device certification. We achieve this by combining the procedures for the estimation of bounds on fidelity with the resulting density matrix obtained by using Choi-Jamio{\l}kowski isomorphism \cite{jamiolkowski1972linear, jamiokowski1974effective, choi1975completely}. We will use the standard representation of this isomorphism. Namely, for a quantum channel $\Phi$, being a completely-positive trace preserving operation on the space of density matrices $\Omega(\Cplx^n)$, we define the Choi-Jamio{\l}kowski image of $\Phi$, $\Jam{\Phi}$, as
\begin{equation}
\Jam{\Phi} = (\Id\otimes \Phi)\sum_{i=1}^n \ket{i} \otimes \ket{i},
\end{equation}
where $\sum_{i=1}^n\ket{i}\otimes\ket{i}$ is the maximally entangled state on the space $\Cplx^n\otimes\Cplx^n$.

\section{Estimation of the fidelity between quantum devices}\label{sec:fidelity-estimation}

\subsection{Problem statement}

Let us assume that the quantum start-up company has developed a device solving an important optimization problem or producing valuable states utilized by some crucial quantum communication protocol. In this case, it is desirable to provide some testing procedure which would ensure the buyers that the offered device really provides the described functionality. To achieve this, one needs to ensure that the operation realized by the quantum device is indeed the one promised by the seller.

In the general case of channel distinguishability, it is customary to assume that we have for our disposal a set of $N$ quantum devices,
represented by quantum channels $\Psi_1,\Psi_2,\ldots,\Psi_N$. Each of the
devices is given as a black box, \ie\ the Kraus representation of the channels
is unknown. In such a situation, one cannot judge if the input devices are
perfectly distinguishable~\cite{duan2009perfect}.

However, in our situation the task is simpler. All we need to do is to convince a customer that the device we would like to sell implements the operation of some standard or ideal device. From now on, we assume that the ideal (or standard) device $D_{\Phi_0}$ is described by a quantum channel $\Phi_0$, which operates on the space of $n$ qubits. We also deal with the second device, $D_{\Psi}$, which is claimed to be identical with $D_{\Phi_0}$.

What's more, as our start-up is a quantum company, its board decided that the certification procedures deployed by the company should also benefit from the quantum advantage. Such decision yields two benefits. Firstly, it supports the claims concerning the ubiquitous applications of quantum computing. Secondly, it provides an opportunity to develop a unique certification service which can be offered to other quantum start-ups~\cite{mohseni2017commercialize}. 


\subsection{SSFB certification procedure}

The first approach one can propose is based on the SSFB bounds. Such approach was first proposed in \cite{puchala2011experimentally}, where new measures of distance between quantum operations have been introduced. These two quantities are~$C_G$ corresponding to root infidelity,
\begin{equation}
C_G(\Phi_0,\Psi) = \sqrt{1-G(\Jam{\Phi_0}, \Jam{\Psi})},
\end{equation}
and $A_{G^2}$, corresponding to Bures angle,
\begin{equation}
A_{G^2}(\Phi_0,\Psi) = \arccos C_G(\Phi_0,\Psi).
\end{equation}

It can be demonstrated that both $C_G$ and $A_{G^2}$ have the properties required from the measures of difference between ideal and real quantum processes~\cite{gilchrist2005distance,puchala2011experimentally}. What is more important, they both can be measured directly. This is due to the fact that super-fidelity, in the terms of which both quantities are expressed, can be measured in the simple quantum procedure~\cite{miszczak2009sub-super}.

\begin{figure}[htp!]
\centering
\begin{quantikz}[thin lines,scale=1.12]
\lstick{\ket{0}} & \gate{H}    & \ctrl{3}   \gategroup[wires=5,steps=4,style={dotted,rounded corners,fill=black!10,inner xsep=2pt},background, label style={label position=below,yshift=-.5cm}]{\sf SSFB}& \ctrl{4} & \gate{H} & \meter{} & \cw \rstick{$p_{\ketbra{0}{0}}, p_{\ketbra{1}{1}}$}  \\
\lstick[wires=2]{\ket{\phi_+}} & \gate{\id} \gategroup[wires=2,steps=1,style={dashed,rounded corners,fill=blue!35,inner xsep=2pt},background]{} & \swap{2} & \qw      & \qw & \qw \\
                               & \gate{\!\Phi_0\!}                                                                                                     & \qw      & \swap{2} & \qw & \qw\\ 
\lstick[wires=2]{\ket{\phi_+}} & \gate{\id} \gategroup[wires=2,steps=1,style={dashed,rounded corners,fill=blue!35,inner xsep=2pt},background]{}  & \swap{}  & \qw      & \qw & \qw \\
                               & \gate{\Psi}                                                                                                     & \qw      & \swap{} & \qw & \qw
\end{quantikz}

\caption{Quantum procedure for calculating the overlap between quantum channels based on super-fidelity~\cite{miszczak2009sub-super}. The procedure consists of the sub-procedure for creating Choi-Jamio{\l}kowski image of the channel (blue dashed parts) applied twice and the sub-procedure for evaluating the overlap (green dotted part). The resulting probability $p_{\ketbra{0}{0}}$ of finding the top qubit in state $\ket{0}$ corresponds to $\tr\Jam{\Phi}\Jam{\Psi} = 2 p_{\ketbra{0}{0}} - 1$ \cite{ekert2002direct}. This leads to the direct estimation of the bound for the fidelity between channels describing devices $\Phi_0$ and $\Psi$. See \cite{miszczak2009sub-super} for the construction of a quantum circuit for measuring sub-fidelity.}
\label{fig:ssfb-measure-circ}
\end{figure}
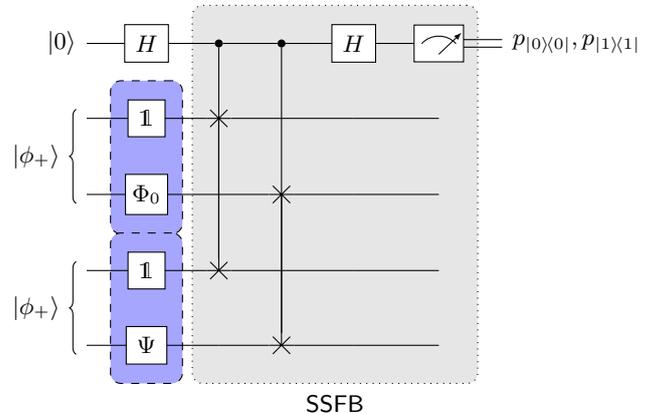

For the sake of completeness, in \figref{fig:ssfb-measure-circ} we recall the construction of the quantum circuit for calculating the overlap between two quantum channels as introduced in~\cite{puchala2011experimentally}.

The circuit presented in \figref{fig:ssfb-measure-circ} consists of two parts. The first part is used to construct the density matrix corresponding to the quantum channel as defined by the Choi-Jamio{\l}kowski isomorphism. The circuit is based on the gate-teleportation approach. The second part of the circuit implements the calculation of the overlap between two density matrices \cite{ekert2002direct}. In this case, as the density matrices fed into the second part of the procedure represent quantum channels, the final measurement results in the estimation of the overlap between the images of the Choi-Jamio{\l}kowski isomorphism for the channels.

\begin{figure}[tbh!]
\centering

\begin{tikzpicture}[auto, node distance=2cm,>=latex',scale=1.12]

\node at (0, 1) [qdev, name=qdev1] {$D_{\Phi_0}$};
\node at (0, -1) [qdev, name=qdev2] {$D_{\Psi}$};

\node at (1.5, 1) [jam1, name=jamPhi0] {$\mathcal{J}$};
\node at (1.5, -1) [jam1, name=jamPsi] {$\mathcal{J}$};

\node at (3.5,0) [ssfb, name=ssfb] {\sf SSFB};

\node [cpu, name=cpu-pp, right=0.75cm of ssfb] {};

\draw [draw,->] (qdev1.east) -- (jamPhi0);
\draw [draw,->] (qdev2.east) -- (jamPsi);

\draw [draw,->] (jamPhi0) -- node[anchor=south,above]{$\Jam{\Phi_0}$}  (ssfb.130);
\draw [draw,->] (jamPsi) -- node[anchor=south,above]{$\Jam{\Psi}$}  (ssfb.-130);

\draw [draw,->] (ssfb.east)+(0.1,0) -- (cpu-pp.west);

\end{tikzpicture}

\caption{SSFB quantum device certification scheme. In this scheme, the goal is to certify a quantum device $\Psi$ against the ideal (or standard) device $\Phi_0$ using the calculable bounds for the fidelity. The input of the procedure is given in the form of physical devices, $D_{\Phi_0}$ and $D_{\Psi}$, implementing channels $\Phi_0$ and $\Psi$, respectively. CPU part of the scheme is responsible for classical post-processing of the results obtained using the quantum procedure. Colours and line styles are identical to the ones used in \figref{fig:ssfb-measure-circ}.}
\label{fig:ssfb-certification-scheme}
\end{figure}
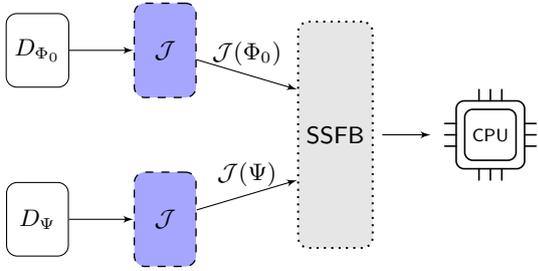

In order to utilize this circuit in the device certification scenario one has to process the obtained measurement results in order to calculate the appropriate bound on the similarity between quantum devices. The diagram representing this procedure is provided in \figref{fig:ssfb-certification-scheme}. The procedure takes two device as an input -- the standard device with the operational capacity already confirmed, and the device for which its conformation with the standard device is to be confirmed.

One should note that in this scheme the classical data processing is required only at the final step of the procedure. This step is required to compute the bounds for the fidelity based on the measurement results.

\subsection{VQFE certification procedure}

The alternative approach to the task of quantum certification can be obtained by utilizing the variational quantum fidelity estimation. The goal of this procedure is to find the truncated fidelity of $\Jam{\Psi}$ for the $m$ largest eigenvalues of $\Jam{\Phi_0}$.

In \cite{larose2019variational}, the procedure for using variational hybrid classical-quantum algorithm was introduced. This procedure can be used to calculate lower and upper bounds for the fidelity based on the \emph{truncated fidelities}, defined as overlaps between two density matrices in the eigenbasis of one of the matrices.

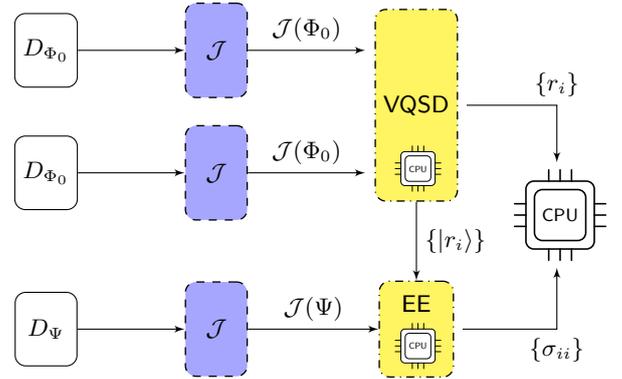
\begin{figure}[htp!]
\centering

\begin{tikzpicture}[auto, node distance=2cm,>=latex',scale=1.12]

\node at (0,  1.75) [qdev, name=qdev1] {$D_{\Phi_0}$};
\node at (0,  0.30) [qdev, name=qdev2] {$D_{\Phi_0}$};
\node at (0, -1.55) [qdev, name=qdev3] {$D_{\Psi}$};

\node at (2,  1.75) [jam1, name=jamPhi01] {$\mathcal{J}$};
\node at (2,  0.30) [jam1, name=jamPhi02] {$\mathcal{J}$};
\node at (2, -1.55) [jam1, name=jamPsi] {$\mathcal{J}$};

\node at (4.35, 1.10) [vqfe, name=vqsd] {\sf VQSD};
\node at (4.35, 0.35) [cpu,scale=0.5] {};

\node at (4.35, -1.55) [vqfe-short, name=vqsd-part] {};
\node at (4.35, -1.25) [] {\sf EE};
\node at (4.35, -1.75) [cpu,scale=0.5] {};

\node at (6, -0.20)[cpu, name=cpu-pp] {};

\draw [draw,->] (qdev1.east) -- (jamPhi01);
\draw [draw,->] (qdev2.east) -- (jamPhi02);
\draw [draw,->] (qdev3.east) -- (jamPsi);

\draw [draw,->] (jamPhi01.east) --node[anchor=south,above]{$\Jam{\Phi_0}$} (3.75,  1.75);
\draw [draw,->] (jamPhi02.east) --node[anchor=south,above]{$\Jam{\Phi_0}$} (3.75,  0.30);
\draw [draw,->] (jamPsi.east) --node[anchor=south,above]{$\Jam{\Psi}$} (vqsd-part);

\draw [draw,->] (vqsd.south)  --node[anchor=south,right]{$\{\ket{r_i}\}$} (vqsd-part.north);

\draw [draw,->] (vqsd.east)+(0.1,0) -| node[anchor=south,above]{$\{r_i\}$}  (cpu-pp.north);

\draw [draw,->] (vqsd-part.east)+(0.1,0) -| node[anchor=west,below]{$\{\sigma_{ii}\}$}  (cpu-pp.south);

\end{tikzpicture}

\caption{VQFE quantum device certification scheme. In this scheme to certify a quantum device $\Psi$ against the ideal (or standard) device $\Phi_0$ the VQFE procedure is used. The input of the procedure is given in the form of physical devices, $D_{\Phi_0}$ and $D_{\Psi}$, implementing channels $\Phi_0$ and $\Psi$, respectively. The first step is to apply Choi-Jamio{\l}kowski isomorphism to obtain dynamical matrices for the input devices. Next, two copies of $\Jam{\Phi_0}$ are used as an input for the variational quantum state diagonalization procedure (VQSD). At the same time, $\Jam{\Psi}$ is processed by a quantum-classical algorithm to obtain its matrix elements in the eigenbasis of $\Jam{\Phi_0}$. Finally, classical processing of the obtained eigenvalues is used to calculate the approximation of the fidelity between quantum operations. Note that in this procedure, CPU part is utilized at several steps -- as a part of VQSD used for calculating eigenvalues $r_i$ of $\Jam{\Phi_0}$ and matrix elements $\sigma_{ij}$ of $\Jam{\Psi}$. Yellow colour indicates hybrid quantum-classical subprocedures. Procedure denoted by {\sf EE}  (Eigenvalue Estimation) evaluates $\sigma_{ij}=\langle r_i|\sigma|r_j\rangle$.} 
\label{fig:vqfe-certification-scheme}
\end{figure}

In the application to the problem of device certification, the quantum variational fidelity estimation consists of the following steps (see \figref{fig:vqfe-certification-scheme}).
\begin{itemize}
\item Prepare two copies of $\Jam{\Phi_0}$ and one copy of $\Jam{\Psi}$.

\item Utilize variational quantum state diagonalization (VQSD) for diagonalizing $\Jam{\Phi_0}$. This procedure outputs eigenvalues $\{r_i\}$ of $\Jam{\Phi_0}$, which can be subsequently stored in a classical CPU, and the eigenvectors of $\Jam{\Phi_0}$, used in the next step.

\item Utilize VQSD with input  $\Jam{\Psi}$ and the eigenvectors of $\Jam{\Phi_0}$ obtained from the first VQSD subprocedure to obtain matrix elements $\sigma_{ii}$ of $\Jam{\Psi}$ in the eigenbasis of $\Jam{\Phi_0}$.

\item The resulting matrix elements of $\Jam{\Psi}$ in the eigenbasis of $\Jam{\Phi_0}$ and eigenvalues of $\Jam{\Phi_0}$ are used to calculate truncated fidelity bounds according to Eq.~(\ref{eq:truncated-fidelity-bounds}).
\end{itemize}

In the remaining part of the paper, in order to simplify the notation, we have considered $\rho = \Jam{\Phi_0}$ and $\sigma = \Jam{\Psi}$.

\section{Comparison of the procedures}\label{sec:proc-compare}

The discussed procedures can be potentially used in the certification scenarios. Naturally, in such case one can take into account various  requirements imposed on the quality or on the cost of the certification. If the figure of merit in the certification scenario is the precision of the procedure, the most important factor is the accuracy of the results given by the procedure. On the other hand, one might be interested in the fast and cost-effective certification of the quantum devices. In this case, the goal is to minimize the size of the systems used during the procedures and the depth of the auxiliary circuits. One should note that in both cases we assume that quantum resources -- the complication of a quantum system, number of copies of quantum states, or the final accuracy -- are more important than the classical resources (eg. time required for classical computation) required for the certification procedure.

\subsection{Memory requirements}

Let us first focus on the requirements concerning the size of quantum systems needed in the case of both discussed procedures. We assume that our devices -- both the standard one and the certified one -- operate on $n$ qubits. The Choi-Jamio{\l}kowski isomorphism leads to the requirement of $2n$ qubits for constructing corresponding dynamical matrices. In the case of SSFB certification, super-fidelity calculation utilizes $4n+1$ qubits as it requires only the values of expressions of the form $\tr\left(\Jam{\Psi_i}\Jam{\Phi_0}\right)$. On the other hand, the quntum circuit for sub-fidelity requires $8n+1$ qubits. Moreover, the sub-fidelity bound requires two copies of both devices to be delivered as an input to the certification procedure. This makes this procedure hard to implement in practice as even if the copies of the standard devices might be available, it is reasonable to assume that each new device undergoing the certification procedure is different.

In the case of VQFE scheme, also two copies of the device $D_{\Phi_0}$ used as the certification model have to used. This is due to the fact that two copies of $\Jam{\Phi_0}$ have to be used by the first VQSD procedure. However, the VQFE certification procedure requires quantum register with $4n+1$ qubits only.

As one can see, the main obstacle for using both procedures is the requirement of using two copies of the dynamical matrix of the standard device. This further translates into the requirement that one has to posses two copies of this device. Moreover, both copies have to be certified to represent the same operation. This problem does not occur in the case of SSFB certification if one is only interested in the upper bound for the fidelity as in this case only one copy of the standard device is required.

It should be also noted that VQFE procedure requires two copies of the standard device to be used only once. If the first VQSD procedure is executed, corresponding eigenvalues and eigenvectors can be utilized to certify newly produced quantum devices. Together with the modest requirements concerning the size of quantum registers, this fact makes VQFE certification procedure more appropriate for practical applications offered to wider audiences.

\subsection{Accuracy of the procedures}

The second aspect of both considered procedures is their accuracy.

First of all, VQFE was developed as a low-rank approximation of the fidelity. In a similar manner, SSFB provides tighter bounds if the input density matrices are low-rank. Thus both certification procedures are better suited for devices operating on low-rank matrices. This translates to the operations with smaller number of Kraus operators~\cite{miszczak2011singular}.

On the other hand, one should note that the bounds obtained by VQFE certification can be tightened monotonically with the rank of the approximation used. Thus, the accuracy of the certification can be refined to achieve the required threshold. SSFB certification does not provide such mechanisms.

To investigate the behaviour of the SSFB and VQFE certification procedures we provide the analysis of the bounds obtained by both methods for the case of random quantum operations~\cite{bruzda2000operations,kukulski2020generating}.  Sampling over the set of quantum operations has potential applications in quantum computing and quantumm information theory, in particular for investigating the properties of quantum circuits~\cite{lu2020direct,sim2020user-specified,thinh2019practical}, analysing data from randomized experiments \cite{granade2017QInfer}, and process tomography~\cite{knee2018quantum}.

\begin{figure}[ht!]
	\centering
	
	\subfigure[$\text{rank}=6$]{
		\includegraphics[]{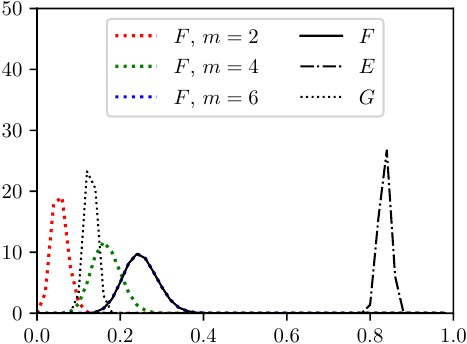}
		\label{fig:bounds-random-operations-rank-6}
	}
	\subfigure[$\text{rank}=10$]{
		\includegraphics[]{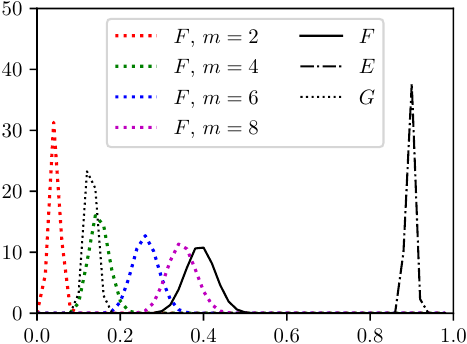}
		\label{fig:bounds-random-operations-rank-10}
	}
	
	\caption{Probability density of values of the bounds for random quantum operations obtained using SSFB and VQFE certification. In the above figure random quantum operators were sampled using the method described in \cite{bruzda2000operations} and implemented in \cite{miszczak2020qi}. A sample of $10^5$ random operations acting on two qubits was used, with dynamical matrices of \subref{fig:bounds-random-operations-rank-6} rank $6$ and \subref{fig:bounds-random-operations-rank-10} rank $10$. In the case \subref{fig:bounds-random-operations-rank-6} truncated fidelity for $m=6$ corresponds to fidelity. Black dotted and black dashdotted lines indicate the values of sub-fidelity $G$ and super-fidelity $E$ obtained for this case. Solid black lines indicate the values of~$F$.
	}
	\label{fig:bounds-random-operations}
\end{figure}

For the purpose of the presented study we utilize the implementation of the random operations generation procedure provided in \cite{miszczak2020qi, miszczak2012generating,miszczak2013employing}. To sample the space of quantum operations we generate a square matrix $X$ of $N^2\times M$ from the complex Ginibre ensemble~\cite{ginibre1965statistical} with elements being complex random Gaussian variables. By taking the partial trace of matrix $GG^\dagger$ over the first subsystem and subsequently normalizing the results we obtain a random dynamical matrix,
\begin{equation}
\left(\id_N\otimes \frac{1}{\sqrt{\tr_N GG^\dagger}}\right)GG^\dagger\left(\id_N\otimes \frac{1}{\sqrt{\tr_N GG^\dagger}}\right)
%
%
\end{equation}
where ${\tr_N}$ denotes the partial trace over the first $N$-dimensions.

\begin{figure}[t!]
	\centering   
	\subfigure[\texttt{CZ} as entangler.]{\label{fig:5a}\includegraphics[width=85mm]{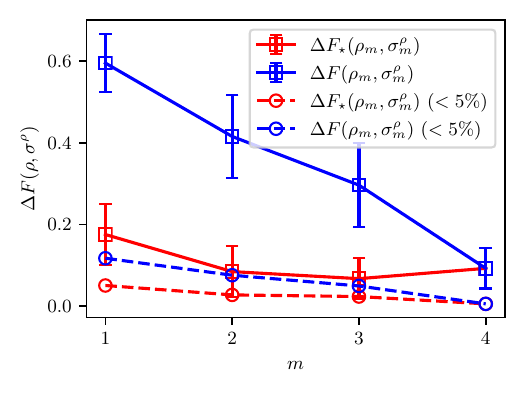}}
	\subfigure[\texttt{CX} as entangler.]{\label{fig:5b}\includegraphics[width=85mm]{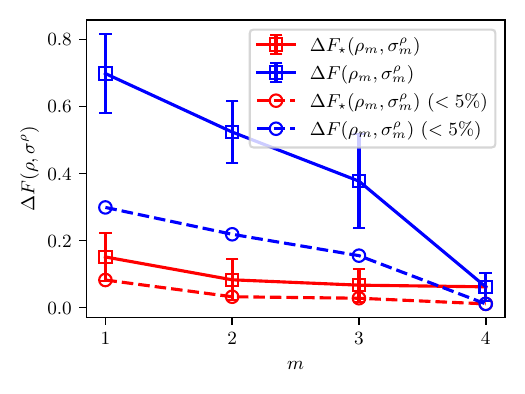}}
	\caption{Error in fidelity estimation with respect to $m$. The results were obtained by taking an average over $1000$ random 1-qubit quantum channels ($n = 1$) of rank~$4$, using IBM quantum computer simulator from 
		\texttt{Qiskit Aer} package. Here $\Delta F(\rho, \sigma^\rho) = F(\rho_m, \sigma_m^\rho) - F(\rho, \sigma)$. The dashed lines depict the average over channels with less than $5\%$ error in the estimation of fidelity. The source code for the implementation can be obtained from~\cite{kundu2021qiskit}.}
	\label{fig:qbit1_rank4_truncated_fidelity_boun}
\end{figure}

The results obtained for the sample of random quantum operations are presented in \figref{fig:bounds-random-operations}. As one can see, to obtain a viable approximation of fidelity in terms of truncated fidelities, the VQFE certification procedure needs to be used to estimate almost all eigenvalues of the dynamical matrix. In this particular case, the lower bound obtained using SSFB provides an estimation similar to the VQFE certification with $m=4$ eigenvalues.

From the results for random quantum operations one can see that VQFE provides a very good tool for lower-rank operations. At the same time SSFB, certification can provide useful lower bound for fidelity between operations. However, the upper bound obtained in SSFB case is unsuitable for providing a viable approximation of fidelity.

\begin{figure}[t!]
\centering
\includegraphics[]{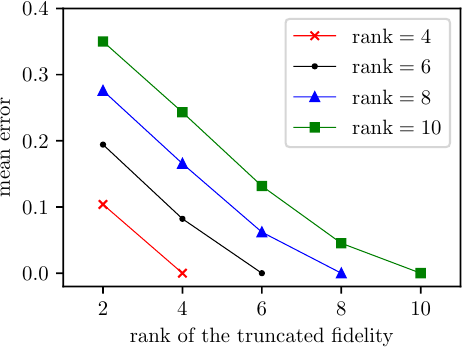}
\caption{Mean error for the approximation of fidelity by the truncated fidelities. Values are plotted for random quantum operations as a function of the rank of the truncated fidelity. Each combination of colour and shape  corresponds to density matrices with a fixed rank. Each point was obtained by by averaging the difference between the fidelity and the truncated fidelity on the sample of $10^5$ pairs of random dynamical matrices.}
\label{fig:truncated-fid-error}
\end{figure}

It can be also noted that the approximation given by VQFE procedure can be significantly improved by increasing the number of eigenvalues estimated in the variational diagonalization procedure. This can be seen in \figref{fig:truncated-fid-error}, where the dependency of the mean approximation error of the truncated fidelity is plotted for dynamical matrices with different rank. As one can see, the bound obtained using truncated fidelities can be easily tightened. Moreover, the mean for the given rank of the truncated fidelity decreases with the increasing rank of random dynamical matrices.

Fidelity estimation with respect to the number of estimated eigenvalues can be observed in \figref{fig:qbit1_rank4_truncated_fidelity_boun}. As an ansatz we used three layers of the following structure
\begin{equation}	
U(\vec{\theta}) = U_\text{ent}\times\prod_{i=0}^{N-1}\texttt{RY}(\theta_i)^{\otimes i} \texttt{RZ}(\theta_i)^{\otimes i},\label{eq:ansatz}
\end{equation}
where $n$ is the dimension of quantum channel. In subfigure (\ref{fig:qbit1_rank4_truncated_fidelity_boun}a) we use $U_\text{ent} = \texttt{CZ}$ and in subfigure (\ref{fig:qbit1_rank4_truncated_fidelity_boun}b) setting $U_\text{ent} = \texttt{CX}$ and $N=2n = 2$.
For simulation purposes we utilize \texttt{Qiskit} package to implement the variational fidelity estimation for one-qubit quantum channels, which can be executed using a quantum computer. One can observe that the difference between the real and the approximated value decreases with the rank of the approximation.

\begin{figure}[t!]
	\centering   
	\includegraphics[width=85mm]{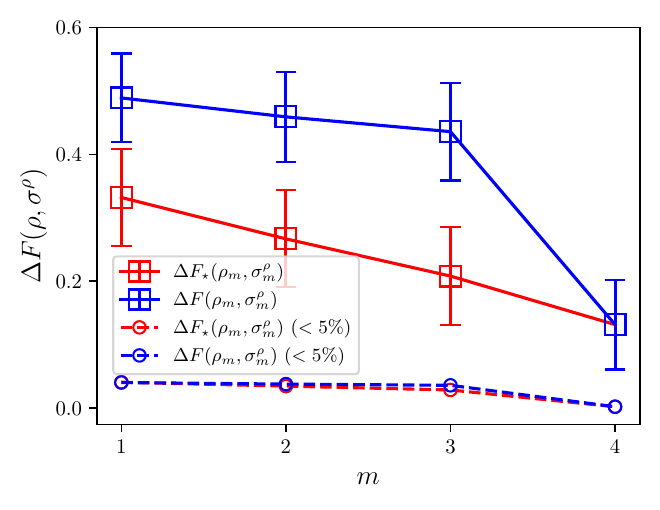}
	\caption{The average error in fidelity estimation with respect to $m$, obtained by averaging over $1000$ random 2-qubit quantum channels ($n = 2$) of rank~$4$, using IBM quantum computer simulator from \texttt{Qiskit Aer} package.}
	\label{fig:qbit2_rank4_truncated_fidelity_boun}
\end{figure}

In the \figref{fig:qbit2_rank4_truncated_fidelity_boun} we illustrate the error in the fidelity estimation for two-qubit quantum channels of rank $4$. It should be noted that although the average error over $1000$ random one and two qubit channels is almost similar in both cases, the difference lies in the number of channels with less than $5\%$ error. While for one qubit channels we get $480$ channels with less than $5\%$ error (dashed lines in \figref{fig:qbit1_rank4_truncated_fidelity_boun}), for two qubit channels the number decreases almost $5$ times i.e. we note $99$ channels with less than $5\%$ error (dashed lines in \figref{fig:qbit2_rank4_truncated_fidelity_boun}). To obtain the results we utilize the similar structure of the ansatz as in Eq.~(\ref{eq:ansatz}) with $N=2n=4$. The entangler is of the following form 
\begin{equation}	
	U_\text{ent} = \prod_{i=0}^{N-1}\texttt{CX}_{i,i+1 (\mathrm{mod} N)}.
\end{equation}

Interestingly, during the simulation we found out that some of the channels are easier to certify than others, which is reflected by the statistics over the number of channels that are converging to the true fidelity for one and two qubit channels, as described above.
\subsection{Sensitivity to noise}
\begin{figure*}[t!]
	\centering
	\includegraphics[width=1.8\columnwidth]{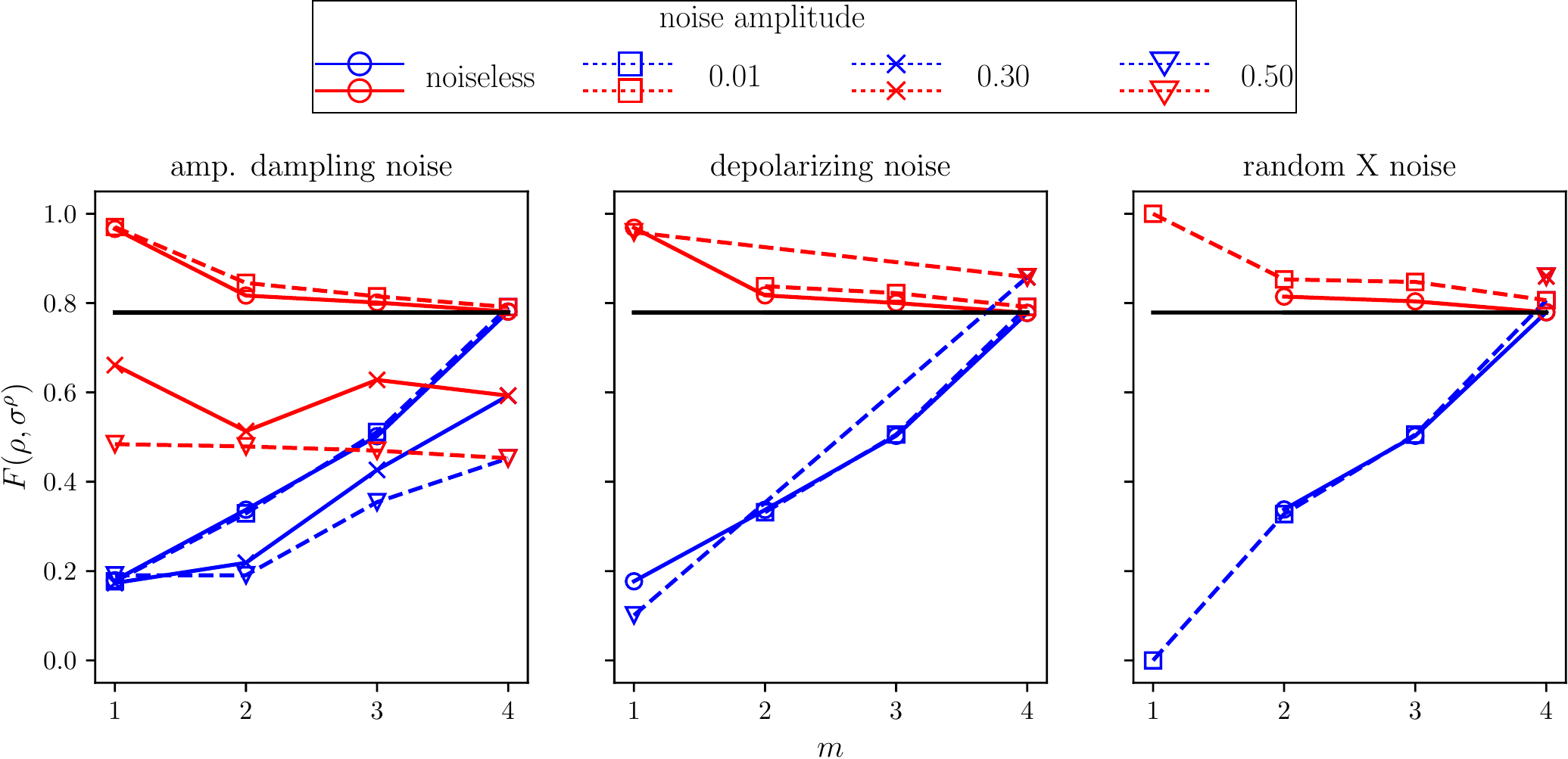}
	\caption{Effect of noise in the estimation of truncated fidelity bound under amplitude damping, depolarizing, and random~$X$ channels. It can be seen that the effect of amplitude damping noise on fidelity estimation is more prominent when the noise ratio exceeds $30\%$. Additionally, for amplitude damping noise when the noise amplitude is within $10\%$, we do not observe any notable changes in the fidelity estimation.}
	\label{fig:noise-model-on-fidelity-estimation}
\end{figure*}
The last aspect of importance is the sensitivity of the proposed procedure. In this section we study the convergence of truncated fidelity bounds under a class of noise models. Let us start by discussing briefly the nature of the noise models under consideration. A $p$-dimensional depolarizing channel can be viewed as CPTP map, which maps a state $\rho$ into linear combination of a maximally mixed state and the state itself
\begin{equation}
	\Delta_\gamma = (1-\gamma)\rho + \frac{\gamma}{p}\id.
\end{equation}
On the other hand, amplitude-damping channel leads to a decay of energy from an excited state to the ground state depending on the probability $\gamma$. Hence, the channel's action on a state is given as 
\begin{equation}
	A_\gamma = \mathcal{K}_0\rho\mathcal{K}_0^\dagger + \mathcal{K}_1\rho\mathcal{K}_1^\dagger,
\end{equation}
where 
\begin{equation}
\mathcal{K}_0 = 
	\begin{bmatrix}
	1 & 0 \\
	0 & \sqrt{1-\gamma} 
	\end{bmatrix},\;\;
\mathcal{K}_1 = 
\begin{bmatrix}
0 & \sqrt{\gamma}\\
0 & 0
\end{bmatrix}.
\end{equation}
Meanwhile, random $X$ quantum channel is defined as
\begin{equation}
	R_\gamma = \gamma X + (1-\gamma)\id,
\end{equation}
where  we apply $X$-gate with probability $\gamma$. In \figref{fig:noise-model-on-fidelity-estimation} we illustrate the effect of depolarizing, amplitude damping, and random $X$ noise. It can be observed that the fidelity bound fails to converge to the true value as the noise amplitude increases by more than $10\%$. The depolarizing and random $X$ noise increases the fidelity at full rank, while due to the effect of amplitude damping noise the estimated fidelity decreases. When the amplitude of noise increases by more than $10\%$, the deviation of fidelity estimation from its true value becomes more prominent.
\begin{figure*}[t!]
	\centering  
	\subfigure[CZ as entangler]{\label{fig:9a}\includegraphics[width=80mm]{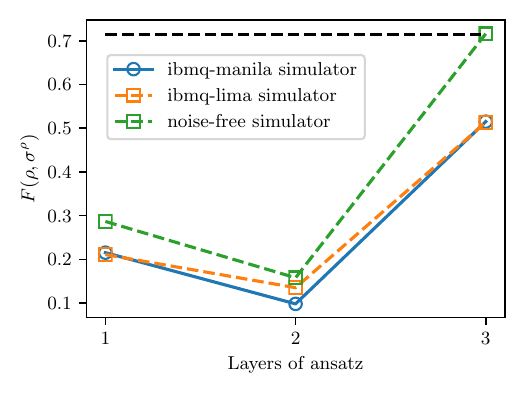}}
	\subfigure[CX as entangler]{\label{fig:9b}\includegraphics[width=80mm]{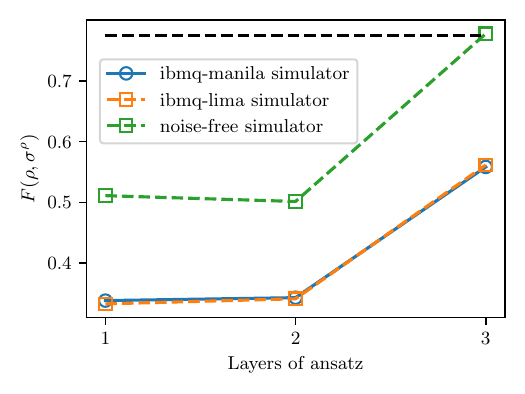}}
	\caption{An example of the convergence of the truncated fidelity for a single random one-qubit (rank $4$) channel in IBM's noise-free simulator and \texttt{AerSimulator} from Qiskit package with real device backends. The black dashed line depicts the true value of fidelity. See~\cite{kundu2021qiskit} for the implementation details.}
	\label{fig:real-device-simulate}
\end{figure*}

To get some insight into the sensitivity of the described procedure in the realistic model, one can consider a noise model of the quantum hardware. To achieve this we run the procedure on IBM quantum computer with noisy and noise-free simulator for the sample of random quantum operations. 

In \figref{fig:real-device-simulate}, we depict the results for fidelity estimation after running the algorithm in \texttt{Qiskit Aer} with real device backends provided by IBM.

\section{Final remarks}\label{sec:final}

As suggested in \cite{cerezo2020variational}, variational quantum fidelity estimation can be naturally used to compute the distance between quantum operations and thus provide means for the certification of quantum devices. 
Following this direction, we have proposed a method for certifying quantum devices which is aimed to be used on near-term intermediate-scale quantum computers. The proposed scheme is particularly suitable for the cases when one needs to provide a convincing argument supporting their claim concerning some quantum mechanical device, and the certified quantum device operates on a relatively small number of qubits. This is exactly the case for the NISQ generation of quantum devices, which makes the presented approach especially appealing in this scenario.

The benefits and features of the VQFE certification procedure are related to the very principles of VQE computing paradigm~\cite{moll2018quantum}. In particular, VQE is one of the main candidates for exploiting the advantage offered by early quantum devices. At the same time, VQE is, by design, suitable for harnessing the strengths of a given architecture. For example, if some gates or quantum operations may be performed with higher fidelity then by utilizing proper ansatz, VQE can benefit from this fact. At the same time, VQE is able to suppress some kinds of quantum errors~\cite{mcclean2016theory}, which makes it especially attractive for NISQ devices. Meanwhile, from the noisy simulation of the certification we can infer that the effect of amplitude damping noise is far more prominent than the depolarizing and random X noise when the noise amplitude is more than $0.1$.

The main weakness of the approach presented in this work is that both procedures require the standard device to be available and it has to be confirmed as operating in accordance to the specification. Thus, one needs some other form of the certification in order to construct the standard device. However, the proposed procedures enable the fast certification based on the quantum resources. Even if the certification of the device used as the standard one is expensive in terms of computational resources, the quantum mechanical approach to the certification makes the certification of next devices significantly cheaper.

This weakness is even more visible in the case of SSFB certification as two copies of the standard device are required. This translates to the requirement of preparing two images of Choi-Jamio{\l}kowski isomorphism corresponding to the standard devices. Equivalently one has to posses two copies of such device and to certify that they represent the same operation. The variational state diagonalization routine used in VQFE as a subprocedure also requires two copies of the device. However, in this case one has to use hybrid quantum-classical procedure for variational diagonalization of the standard device only once. This fact and the quality of the obtained certification makes the VQFE certification an attractive approach to the problem of quantum device certification, especially when a significant number of new devices must undergo the certification procedure.

Eigen-decomposition of Choi-Jamio{\l}kowski states is an indespensible recipe to obtain the Kraus operators of an unknown quantum operation \cite{leung2003choi}. The VQSD algorithm along with Choi-Jamio{\l}kowski isomorphism provides complete eigen-information of an unknown quantum operation hence the diagonalizing subroutine of the certification process can be utilized in quantum process tomography. The introduced scheme is based on the well-known and studied building blocks used in the theory of quantum computing. We believe that this demonstrates the current trend in the area of quantum data processing when one aims at building practical applications of quantum computing by utilizing the concepts and methods which were quite recently of only theoretical interest. The presented work demonstrates that such approach can lead to interesting applications of quantum computers.


\acknowledgments
This work has been partially supported by the Polish National Science Center (NCN) under the grant agreement 2019/33/B/ST6/02011.

JM would like to express his gratitude to P.~Gawron, Z. Pucha{\l}a, and K. \.Zyczkowski for discussions concerning the problems related to process discrimination and generation of random quantum operations. Authors would also like to acknowledge comments from  M.~Ostaszewski concerning the variational quantum computing and help from A. Glos with improving the implementation of the variational fidelity estimation.

\bibliographystyle{plainnat}
\bibliography{compare_quantum_devices.bib}

\end{document}